\documentclass[aps,pre,twocolumn,floatfix,nofootinbib,showpacs,groupedaddress,superscriptaddress]{revtex4}
\usepackage[sort&compress]{natbib}
\usepackage{graphicx}
\usepackage{epsfig}

\begin{document}
\title{Protein-mediated Loops and Phase Transition in Nonthermal Denaturation of DNA}
\author{K.G. Petrosyan}
\affiliation{Institute of Physics, Academia Sinica, Nankang, Taipei
11529, Taiwan}
\author{Chin-Kun Hu}
\affiliation{Institute of Physics, Academia Sinica, Nankang,
Taipei 11529, Taiwan} \affiliation{Center for Nonlinear and
Complex Systems and Department of Physics, Chung-Yuan Christian
University, Chungli 32023,  Taiwan}
\date{\today}

\begin{abstract}
We use a statistical mechanical model to study nonthermal
denaturation of DNA in the presence of protein-mediated loops. We
find that looping proteins which randomly link DNA bases located at
a distance along the chain could cause a first-order phase
transition. We estimate the denaturation transition time near the
phase transition, which can be compared with experimental data. The
model describes the formation of multiple loops via dynamical
(fluctuational) linking between looping proteins, that is essential
in many cellular biological processes.
\end{abstract}


\maketitle

Denaturation of DNA is a fundamental biological process before the
transcription stage \cite{watson}. Thermal denaturation of DNA
\cite{wartell} has been modelled in many ways, including the
ladder \cite{lavis}, Poland-Scheraga \cite{scheraga} and
Peyrard-Bishop \cite{peyrard} models. The process still attracts
attention of theoreticians in an attempt to describe it most
efficiently \cite{weber}. Besides the melting, DNA also denatures
under the influence of other factors such as pH value, salt
concentration, other chemical factors, and mechanical forces. One
example of the latter is the DNA denaturation induced by an
externally applied torque. The experiments with single DNA
molecules under torsional stress were reported in \cite{strick,
bryant} that shed more light on the mechanical properties of DNA
molecules in connection with their functioning in living cells.
{\it In vivo} the torque is exerted by the RNA polymerase that
causes transcription-generated torsional stress \cite{kouzine}
(see also \cite{harada} where a direct observation of DNA rotation
during transcription by {\it Escherichia coli} RNA polymerase was
reported). A theoretical study of torque-induced DNA denaturation
was presented by Cocco and Monasson \cite{cocco} and a thorough
investigation of the effect of mechanical forces and torques on
DNA and its denaturation was done by Marko \cite{marko}.

Here we are interested in {\it nonthermal} denaturation of DNA
that precedes the transcription process. Transcription regulation
typically involves the binding of proteins over long distances on
multiple DNA sites which are then brought close to each other to
form DNA loops \cite{saiz}. The DNA loops can be formed by protein
complexes, e.g., by the regulators of bacterial operons, such as
ara, gal, and lac, and human proteins involved in cancer, such as
retinoic X receptor. The presence of protein-mediated loops is
also important for many other cellular processes, including DNA
replication, recombination, and nucleosome positioning as was
extensively discussed in \cite{sv}.

Recently Vilar and Saiz \cite{vilar} studied  multiprotein DNA
looping. They developed a model of formation of a single loop via
connection of an arbitrary large number of proteins. Their model
describes a switchlike transition between looped and unlooped
phases, and has been extended to account for multiple loops
\cite{sv}. Dynamic protein-mediated loops within the framework of
molecular systems biology were considered in \cite{saiz} for the
cases of the lac operon and phage $\lambda$ induction switches.
Here we consider a different model to describe the denaturation of
DNA, which has loops formed by proteins that link bases randomly
located along the molecular chain. Thus our model accounts for
formation of multiple loops that is essential in cellular
biological processes like pre-mRNA splicing \cite{watson}. Yet
another important feature of our model is that it presents a
dynamic rather than static picture of formation of loops as the
protein-mediated links between the base pairs fluctuate, {\it
i.e.} the proteins couple and decouple in the course of time. This
demonstrates a connection between formation of the structure of
protein-mediated loops for the particular DNA-protein node-link
interaction network and co-evolutionary complex networks
\cite{statnets, dorogov}. We show below that looping proteins can
make the nonthermal denaturation process to be a first-order phase
transition. It is due to the effective long-range interactions by
the mediating proteins. We are primarily interested in the phase
transition, in the metastability phenomenon that we have found and
in the kinetics of the denaturation. We then calculate the
transition time from the double-helix state to the coil state,
which can be compared with experimental data.

\vskip 2 mm

{\bf The Model.} Lattice models proved to be useful in studies of
the phenomenology of DNA denaturation \cite{palmeri}. Here we
consider a simple statistical mechanical model defined on an
one-dimensional lattice with each site corresponding to a
rung of the ladder \cite{lavis}. A spin variable $\sigma_i$ is
associated with each site $i$ where $\sigma_i=-1$ when the
corresponding $H$-bond is intact and $\sigma_i = +1$ when it is
broken. We assume an arbitrary folding of the DNA molecule so that
any two base pairs may get connected via the looping proteins. The
proposed model has the following Hamiltonian
\begin{eqnarray}
H= &-& g\sum_{i=1}^{N-1}\sigma_i\sigma_{i+1} -\gamma\sum_{i<j}
J_{ij} p_i p_j \sigma_i \sigma_j \nonumber \\
&-& h\sum_{i=1}^N\sigma_i - \epsilon\sum_{i=1}^N p_i +
\alpha\sum_{i<j} J_{ij} \label{dna}
\end{eqnarray}
where $g>0$ is the coupling parameter of nearest-neighbor
interactions; $J_{ij}$ are the link variables, taking values $0$
and $1$ when the $i$ and $j$ nodes are uncoupled or coupled by the
proteins, correspondingly; the absence or presence of proteins at
site $i$ is defined by the variable $p_i$ that takes values 0 or
1, respectively; $\gamma>0$ is the energy of interaction between
the base pair sites coupled via an appeared link caused by the
on-site proteins; $h$ is the binding energy between base pairs
that includes the energy of the hydrogen bonds; $\epsilon$ is the
energy of binding of a protein at the site $i$; $\alpha$ is the
energy of formation of a link connecting $i$ and $j$ sites; $N$ is
the number of base pairs.

The first term assures that broken pairs tend to break pairs next
to them and in the same way it makes to pair up bases next to
paired ones. The second term describes creation of links between
proteins bound to bases at random sites of the molecule. These
links form protein-mediated loops. The links actually fluctuate as
the proteins at different sites may couple and decouple in the
course of time. In general, $\gamma$ may depend on the length of
the loop. However, such a dependence is a higher order effect and
we do not consider that. In the third term the energy $h$ depends on
the external parameters that are determined by environmental
conditions such as temperature $T$, pH value, salt concentration,
and other chemical as well as mechanical factors. Change in $h$
may cause openings and closings of base pairs. As an example we
will consider its dependence on an externally applied torque. The
energy $h$ is a sum of contributions from the base pairing energy
$h_0<0$ and from the torsional energy $h_{\tau}$, associated with
a change in the local twist, that is $h=h_0+h_{\tau}$, where
$h_{\tau}=(1/2)C(\Delta\omega)^2$, with $C$ being the twisting
elastic constant (torsional stiffness) and
$\Delta\omega=\omega-\omega_0$ being the deviation of the spatial
angular frequency $\omega$ (change of the rung angle around the
axis per unit length along the chain) from its unstressed value
$\omega_0$ \cite{marko siggia}. The torsional energy can also be
represented via the torque $\tau$ as $h_{\tau}=\tau\varphi_0$,
where $\varphi_0=2\pi/10.5=0.6$ radians per base pair (double
helix contains about 10.5 base pairs per helical turn). The fourth
term  is the energy of binding of proteins. The last (fifth) term is
the energy of formation of a link between base pairs mediated by
the looping proteins. We will use another parameter $c$ defined
via ${c}/{(N-c)} = e^{-\alpha\beta}$, where $\beta$ is the inverse
temperature. The ratio can be roughly treated as the probability
of a link formation (see \cite{statnets} for more rigorous
formulations and details of a related model that describes a
network of fluctuating links). We will assume sparse connectivity
${c}/{N} \ll 1$ with the number of looping proteins much less than
the number of bases. The Hamiltonian may also include
long-range direct H-bond interactions between open base pairs via a term
proportional to $\sum_{ij} A_{ij} (1+\sigma_i)(1+\sigma_j)$
with an interaction matrix $A_{ij}$. However we neglect these interactions
assuming that their contribution is smaller compared to the interactions
mediated by proteins \cite{energy values}.

\vskip 2 mm

{\bf Effective Hamiltonian and Free Energy.} The small number of
proteins compared to the number of base pairs allows us to reduce
(\ref{dna}) to an effective mean-field type Hamiltonian. For that
purpose we eliminate consequently $J_{ij}$ and $p_i$ variables
while calculating the partition function $Z=Tr_{J, p,
\sigma}e^{-\beta H}$, where the trace means summing up over the
corresponding variables. Taking the trace over $J_{ij}$'s
\cite{statnets}
we arrive at the partition function $Z \propto Tr_{p,
\sigma}e^{-\beta H'}$ with the following effective Hamiltonian
\begin{eqnarray}
H' = - g\sum_{i=1}^{N-1}\sigma_i\sigma_{i+1} - \gamma' \sum_{i<j}
p_i p_j \sigma_i \sigma_j \nonumber \\
- \lambda \sum_{i<j}p_i p_j - h\sum_{i=1}^N\sigma_i -
\epsilon\sum_{i=1}^N p_i . \label{effective}
\end{eqnarray}
Here $\gamma' = (c/N)\sinh\beta\gamma$ and $\lambda =
(c/N)(\cosh\beta\gamma - 1)$. The Hamiltonian (\ref{effective})
describes a system consisting of two interacting subsystems, DNA
and proteins. Different time scales and different temperatures for
two subsystems may lead to novel phenomena \cite{anomalous}.
However, here we assume that DNA and proteins are in contact with
the same heat bath at temperature $T$.

\begin{figure}
\includegraphics[width=0.95\linewidth,angle=0]{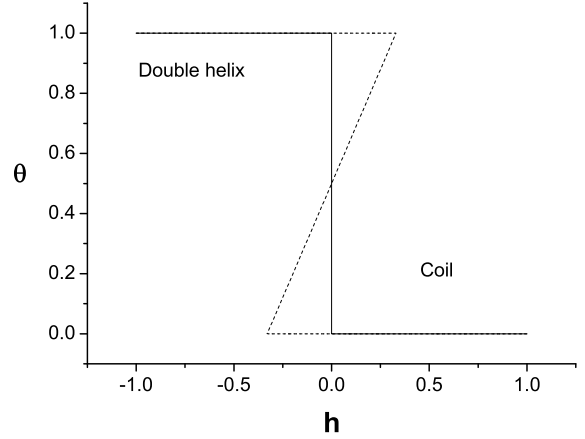}
\hfill \caption{Double helix fraction $\theta = (1-\mu)/{2}$ vs
the base pair binding energy $h$ for the parameters $g = 8.5$
kcal/mol, $\gamma = 0.02$ kcal/mol, $c = 10$, $\epsilon = 7.2$
kcal/mol and $k_B T = 0.6$ kcal/mol. The first-order denaturation
phase transition occurs at the critical value $h_c=0$. The
critical torque is $\tau_c=1.6 k_B T$ for AT-rich and $\tau_c=7
k_B T$ for GC-rich chains. The double helix (coil) becomes
metastable for $h>0$ ($h<0$) as indicated by dashes.} \label{fig1}
\end{figure}

For the case of strong binding energies
$\epsilon\gg\lambda$,$\gamma'$, we can make a mean-field
approximation and replace $p_i$'s by their mean values $\langle p
\rangle = {e^{\beta\epsilon}}/{(1+e^{\beta\epsilon})}$, the
proposed model is then reduced to the following effective
Hamiltonian
\begin{eqnarray}
H_{eff} = - g\sum_{i=1}^{N-1}\sigma_i\sigma_{i+1} -
\gamma''\sum_{i<j} \sigma_i \sigma_j - h\sum_{i=1}^N\sigma_i
\label{Heff}
\end{eqnarray}
where $\gamma'' = \frac{c}{N} \sinh\beta\gamma \cdot
\left(\frac{e^{\beta\epsilon}}{1+e^{\beta\epsilon}}\right)^2$
represents the effective coupling between base pairs mediated by
proteins. Notice that we have neglected the effect caused by the presence
of the persistence length $l_0$ that would require to take into account only
the terms for which one has $|i-j|>l_0$ as the correction would be of order
$O(\frac{l_0}{N})$ and would go to zero in the thermodynamic limit.
The coupling in (\ref{Heff}) is similar to that of a synchronization model
with small world coupling \cite{gade}.

To calculate the partition function $Z \propto Tr_{\sigma}
e^{-\beta H_{eff}}$ for the Hamiltonian (\ref{Heff}), we use the
relationship $\sum_{i<j} \sigma_i \sigma_j = \frac{1}{2}(\sum
\sigma_i)^2 - \frac{1}{2}N$, the Hubbard-Stratonovich
transformation $e^{\frac{1}{2}a(\sum_{i=1}^N
\sigma_i)^2}=\int^{+\infty} _{-\infty} \frac{d \mu}{\sqrt{2\pi/a}}
e^{-\frac{1}{2}a\mu^2 + a\mu\sum_{i=1}^N \sigma_i}$ and the
expression for the partition function of the one-dimensional (1D)
Ising model \cite{lavis}. Then the partition function takes the
form $Z \propto \int^{+\infty} _{-\infty} d\mu e^{- \beta N
f(\mu)}$ with the effective free energy $f(\mu)$ given by
\begin{eqnarray}
f(\mu) &=& \frac{1}{2}b \mu^2 - \beta^{-1}\ln [\cosh \beta (h + b \mu)\nonumber \\
 &+& \sqrt{\sinh^2 \beta (h + b \mu) + e^{-4\beta g}}]. \label{free
energy}
\end{eqnarray}
Here $b =  c \cdot \sinh\beta\gamma \cdot
\left(\frac{e^{\beta\epsilon}}{1+e^{\beta\epsilon}}\right)^2$ and
$\mu$ is the order parameter for the denaturation process. For the
double helix state with all base pairs bound, $\mu=-1$; for the
completely denaturated state, $\mu=1$. The values of $\mu$, which
determine the state of the molecule, are obtained via $f'(\mu)=0$
that leads to the equation
\begin{eqnarray}
\mu = \frac{\sinh\beta (h + b \mu)}{\sqrt{\sinh^{2}\beta (h + b
\mu) + e^{-4\beta g}}}. \label{mean}
\end{eqnarray}

The model is an effective Ising model with 1D nearest-neighbor and
global (all-to-all) interactions. It can be shown that the model
goes through a phase transition provided $\beta b e^{2\beta g}
\geq 1$. That gives the necessary condition for the model
parameters, e.g., the temperature. The sufficient condition for
the phase transition would be the sign change of $h$. Thus $h_c=0$
or $\tau_c=h_0/\varphi_0$ defines the critical point for the
first-order phase transition if necessary condition $\beta b
e^{2\beta g} \geq 1$ is satisfied for the given parameters. The
critical torque $\tau_c$ ranges from $1.6 k_B T$ for weakly bound
(AT-rich) sequences to $7 k_B T$ for the most strongly bound
(GC-rich) sequences \cite{torque values}.

\begin{figure}
\includegraphics[width=0.95\linewidth,angle=0]{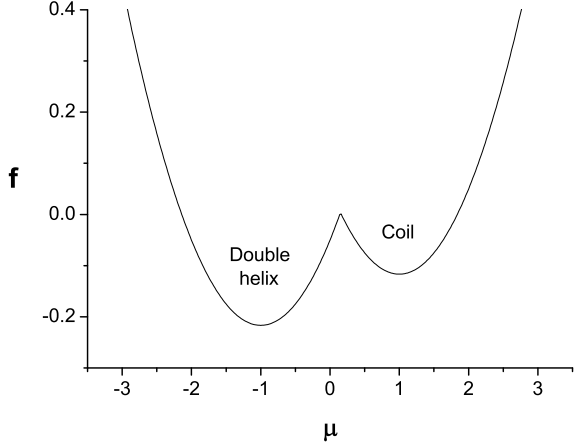}
\hfill \caption{Free energy vs the order parameter $\mu$ at the
base pair binding energy value $h = -0.05$kcal/mol for the
parameters $g = 8.5$ kcal/mol, $\gamma = 0.02$ kcal/mol, $c = 10$,
$\epsilon = 7.2$ kcal/mol and $k_BT = 0.6$ kcal/mol. Double helix
is stable and coil is metastable for $h<0$.} \label{fig2}
\end{figure}

In order to quantify the degree of denaturation we introduce the
parameter $\theta = (1-\mu)/2$ that is the fraction of bound base
pairs. The parameter takes the value $\theta=1$ for the double
helix state and the value $\theta=0$ for the denaturated coiled
state. The dependence of the double helix fraction $\theta$ on $h$
is presented in Fig.\ref{fig1}. There is a metastability in a
range of the controlling external parameter $h$. This effect is
illustrated in Fig.\ref{fig2} where the free energy with two
minima is presented. Notice that there is no phase transition if
the proteins do not interact ($\gamma=0$) and thus the
protein-mediated loops are absent. These are the looping proteins
which provide with the long-range interactions that make it
possible to obtain a phase transition for the effectively 1D
lattice model.

\vskip 2 mm

{\bf Transition time.} The kinetics of the denaturation transition
can be treated via the Langevin equation $\dot \mu = -\Gamma\frac{\partial f(\mu)}{\partial \mu} + \xi (t)$,
where $\Gamma$ defines the inverse relaxation time, $\xi (t)$ is
the while noise satisfying the relation
$\langle\xi(t)\xi(t')\rangle = D\delta(t-t')$ with the diffusion
coefficient $D$ determined by the fluctuation-dissipation relation
$D=2\Gamma k_B T$. The corresponding Fokker-Planck equation (FPE)
for the probability distribution function $P(\mu)$ of the order
parameter $\mu$ is $\dot P = \frac{\partial}{\partial \mu}A(\mu)P
+ \frac{1}{2}D\frac{\partial^2}{\partial \mu^2}P$, where $A(\mu) =
-\Gamma\frac{\partial f(\mu)}{\partial \mu}$. Making the
transformations $P \rightarrow Pe^{\frac{f(\mu)}{D}}$, $D
\rightarrow D\Gamma$ and $t \rightarrow t/\Gamma$, we can rewrite
the FPE  as $-\dot P = HP$ with the Hamiltonian $H =
-\frac{1}{2}D\frac{\partial^2}{\partial \mu^2} +\frac{1}{2D}\Phi^2
+ \frac{1}{2}\frac{\partial \Phi}{\partial \mu}$ where
$\Phi(\mu)=-f'(\mu)$. One can exactly solve the FPE to obtain
$P_t(\mu, \mu_0) = |\psi_0(\mu)|^2 +
\frac{\psi_0(\mu)}{\psi_0(\mu_0)} \sum_{n=1}^{\infty}
e^{-\frac{\lambda_n t}{D}}\psi_n(\mu)\psi_n(\mu_0)$, where
$\psi_0(\mu)\propto e^{-f(\mu)/2k_B T}$ and $\mu_0$ is the initial
value. The decay rates $\lambda_n$ and the eigenfunctions
$\psi_n(\mu)$ can be, in principle, derived exactly \cite{junker}.
However, we are not considering here the dynamics of the
probability distribution function. Our goal is to analyze the
dependence of the transition rate on the model parameters, such as
temperature $T$. Therefore we are only interested in the first
eigenvalue given by $\lambda_1 \simeq \frac{D}{\pi}
\sqrt{f''(\mu_{min})|f''(\mu_{max})|} \cdot
e^{-\frac{2}{D}[f(\mu_{max}) - f(\mu_{min})]}$ which governs the
dynamics for long times.

Let us consider the transition from the left minimum of the free
energy in Fig. 2, corresponding to the native double helix state
of DNA, to the right minimum representing denaturated state at the
critical value $h_c$. At this value, that corresponds to the
first-order phase transition point, the free energy is a symmetric
curve with two equal minima and the maximum located at $\mu=0$.
The transition time $\Omega^{-1}$ (the inverse transition
frequency) is twice the time needed to achieve the top of barrier
$\mu_{max}$ from the minimum $\mu_{min}$ which is obtained from
$\lambda_1$
\begin{eqnarray}
\Omega^{-1} \simeq \frac{2\pi}{\Gamma} \frac{e^{\beta[f(\mu_{max})
- f(\mu_{min})]}}{\sqrt{f''(\mu_{min})|f''(\mu_{max})|}}.
\label{time}
\end{eqnarray}
This is a standard expression for the Kramers problem
\cite{gardiner}. However there is a qualitative difference since
the potential $f(\mu)$ itself depends on temperature. The behavior
of the denaturation transition time versus temperature drastically
differs from the conventional Arrhenius case. Although the
transition time first decreases at very low temperatures (frozen
DNA) it begins to increase at high enough (physiological)
temperatures. The reason is that the second derivative present in
the denominator of Eq.(\ref{time}) at the point $\mu_{max}=0$
diverges since $f''(0)=b(1-\beta be^{2\beta g})$ and $\beta
be^{2\beta g} \rightarrow 1$ ($f''(0) \rightarrow 0$) at the
critical temperature defined by the above mentioned necessary
condition of denaturation. For the set of parameters given in
Figs. 1 and 2, the transition time is $\Omega^{-1}=2.35 \cdot
10^{-5} \Gamma^{-1}$. The kinetics of pH-driven denaturation of
DNA was studied experimentally in \cite{ageno}, where the
transition time for single molecule denaturation was estimated to
be of order of $1 \div 10$ seconds. Taken these values we come up
with the inverse relaxation time $\Gamma$ to be of order of
$10^{-6}$ Hz. However we believe that modern measurements in
experiments with single molecules are needed to find precise
values of the quantities.

\vskip 2 mm

{\bf Discussion.}  In summary, we have introduced and studied a
model of nonthermal denaturation of DNA that can be induced by
chemical factors, such as pH value or salt concentration, or by
externally applied mechanical forces and torques (as an example we
considered the case of torque-induced denaturation) in the
presence of protein-mediated loops. The model accounts for
proteins that bind to the DNA molecule. The bound proteins are
then allowed to interact in a random way with each other thus
creating the loops. We have found a first-order denaturation phase
transition that is caused by the looping proteins, the proteins
that connect base pairs that are at a distance  along the chain.
The model possesses a metastability region provided that the
necessary and sufficient conditions are satisfied. The kinetics of
the denaturation phase transition was described by a stochastic
dynamics for the order parameter that is, in principle, exactly
solvable. However we have been mainly interested in obtaining the
transition rate in the vicinity of the first-order phase
transition. It has the standard form by Kramers with the
associated potential being temperature-dependent. This leads to
deviation from the Arrhenius law at physiological temperatures. In
particular, the transition time becomes extremely large when the
temperature approaches its critical value that is defined by the
necessary condition for the denaturation phase transition. The DNA
denaturation kinetics considered here can be extended spatially to
describe a front propagation process in the presence of
protein-mediated loops and the noise that corresponds to the {\it
in vivo} case.

Finally, we have revealed a new purpose of the protein-mediated
looping that is to facilitate {\it in vivo} denaturation of DNA
needed to take it to the next transcription stage. The model also
describes the formation of multiple loops via dynamical
(fluctuational) linking between looping proteins, that is
essential in cellular biological processes such as the pre-mRNA
splicing \cite{wang cooper} and the phenomenon of genomic
plasticity \cite{alberts}. It can mimick, e.g., the coevolutionary
networks of splicing cis-regulatory elements \cite{xiao} having
the loops to splice out introns thus defining the exons within the
DNA molecule. The presented theory can be applied in studies of
the above enumerated {\it in vivo} processes as well as for
description of {\it in vitro} experiments with single DNA molecules.
Yet another application of this or a generalized statistical mechanics model
would be an investigation of dynamic genome architecture
in eukaryotic cells \cite{nicodemi}.

We thank A.E. Allahverdyan, D. Mukamel, E.I. Shakhnovich, and M.C. Williams
for comments and discussions. This work was supported by Grants NSC
96-2911-M 001-003-MY3, NSC 96-2811-M 001-018, NSC
97-2811-M-001-055 \& AS-95-TP-A07, and by the National Center for
Theoretical Sciences in Taiwan.


\begin{thebibliography}{99}
\bibitem{watson}J.D. Watson et al, {\it Molecular Biology of the Gene} (Benjamin/Cummings,
Menlo Park, Calif., 2004).

\bibitem{wartell}R.M. Wartell and A.S. Benight, Phys. Rep. {\bf 126}, 67 (1985).

\bibitem{lavis}D.A. Lavis and G.M. Bell, {\it Statistical mechanics of lattice
systems}, v. 1, p. 57 (Springer-Verlag, Berlin, 1999).

\bibitem{scheraga}D. Poland and H.A. Scheraga, J. Chem. Phys. {\bf 45},
1456 (1966); {\it ibid}, 1464 (1966).

\bibitem{peyrard}M. Peyrard and A.R. Bishop, Phys. Rev. Lett. {\bf 62}, 2755 (1989).

\bibitem{weber}G. Weber et al, Nature Phys. {\bf 2}, 55 (2006).

\bibitem{strick}T.R. Strick et al, Biophys. J. {\bf 74}, 2016 (1998); Physica {\bf 263A}, 392 (1999);
J.F. L\'{e}ger et al, Phys. Rev. Lett. {\bf 83}, 1066 (1999).

\bibitem{bryant}Z. Bryant et al, Nature {\bf 424}, 338 (2003); J. Gore et al, Nature (2006); M.
N\"{o}llmann et al, Nat. Struct. Mol. Biol. {\bf 14}, 264 (2007).

\bibitem{kouzine}F. Kouzine et al, Nat. Struct. Mol. Biol. {\bf 11}, 1092 (2004).

\bibitem{harada}Y. Harada et al, Nature {\bf 409}, 113 (2001).

\bibitem{cocco}S. Cocco and R. Monasson, Phys. Rev. Lett. {\bf 83}, 5178 (1999).

\bibitem{marko}J.F. Marko, Phys. Rev. E {\bf 76}, 021926 (2007).

\bibitem{saiz}J.M.G. Vilar and L. Saiz, Curr. Opin. Genet. Dev. {\bf 15}, 136 (2005);
J.M.G. Vilar and S. Leibler, J. Mol. Biol. {\bf 331}, 981 (2003);
L. Saiz and J.M.G. Vilar, Curr. Opin. Struct. Biol. {\bf 16}, 344
(2006); Nature/EMBO Molecular Systems Biology {\bf 2},
doi:10.1038/msb4100061 (2006); G.A. Maston et al, Annu. Rev. Genomics Hum. Genet. {\bf 7}, 29 (2006).

\bibitem{sv}L. Saiz and J.M.G. Vilar, Nucleic Acids Res. {\bf 36},
726 (2008).

\bibitem{vilar}J.M.G. Vilar and L. Saiz, Phys. Rev. Lett. {\bf 96}, 238103 (2006).

\bibitem{statnets}A.E. Allahverdyan and K.G. Petrosyan, Europhys. Lett. {\bf 75}, 908 (2006).

\bibitem{dorogov}S. N. Dorogovtsev et al, Rev. Mod. Phys. {\bf 80}, 1275 (2008).

\bibitem{palmeri}J. Palmeri et al, Phys. Rev. Lett. {\bf 99}, 088103 (2007); Phys. Rev. E {\bf 77}, 011913
(2008); R. Everaers et al, {\it ibid.} {\bf 75}, 041918 (2007).

\bibitem{marko siggia}J.F. Marko and E.D. Siggia, Phys. Rev. E {\bf 52},
2912 (1995).

\bibitem{energy values}The energy of protein-protein interactions is estimated to be of the
order of $10 \div 20$ kcal/mol \cite{oakeley} compared to the energy of a few kcal/mol for hydrogen
bonds formed between DNA bases \cite{torque values}. Besides, the probability of H-bond formation
during DNA's 3D motion is decreased as the open bases are required to align appropriately
to form the bond.

\bibitem{oakeley}A. Tomovic and E.J. Oakeley, PLoS ONE {\bf 3}, e3243 (2008).

\bibitem{anomalous}A.E. Allahverdyan and K.G. Petrosyan, Phys. Rev. Lett. {\bf 96}, 065701
(2006).

\bibitem{gade}P.M. Gade and C.-K. Hu, Phys. Rev. E {\bf 62}, 6409 (2000).

\bibitem{torque values}P. Thomen et al, Phys. Rev. Lett. {\bf 88},
248102 (2002); T.R. Strick et al, Genetica {\bf 106}, 57 (1999); J.F. Marko, in {\it Multiple
aspects of DNA and RNA: From Biophysics to Bioinformatics}, Les
Houches Session LXXXII, p. 211 (Elsevier,Amsterdam, 2005).

\bibitem{junker}G. Junker, {\it Supersymmetric Methods in Quantum and Statistical
Physics} (Springer-Verlag, Berlin, 1996).

\bibitem{gardiner}C.W. Gardiner, {\it Handbook of Stochastic Methods}
(Springer-Verlag, Berlin, 1985).

\bibitem{ageno}M. Ageno et al, Biophys. J. {\bf 9}, 1281 (1969).

\bibitem{wang cooper}G.-S. Wang and T.A. Cooper, Nature Rev. Genet.
{\bf 8}, 749 (2007) and references therein.

\bibitem{alberts}B. Alberts et al, {\it Molecular Biology of the
Cell}, 4th ed. (Garland Science, New York, 2002).

\bibitem{xiao}X. Xiao et al, PNAS {\bf 104}, 18583 (2007).

\bibitem{nicodemi}M. Nicodemi and A. Prisco, Biophys. J. {\bf 96}, 2168 (2009).

\end{thebibliography}
\end{document}